\documentclass[fleqn,10pt]{wlscirep}
\usepackage[utf8]{inputenc}
\usepackage[T1]{fontenc}
\usepackage{soul}
\usepackage{subfiles}

\title{Creativity in social networks is enhanced by `Goldilocks' dispersal of ideators' visibility}

\author[1]{Raiyan Abdul Baten}
\author[2]{Richard N. Aslin}
\author[3]{Gourab Ghoshal}
\author[4,1,*]{Ehsan Hoque}
\affil[1]{Department of Electrical and Computer Engineering, University of Rochester, NY, USA}
\affil[2]{Haskins Laboratories and Department of Psychology, Yale University, CT, USA}
\affil[3]{Department of Physics and Astronomy, University of Rochester, NY, USA}
\affil[4]{Department of Computer Science, University of Rochester, NY, USA}

\affil[*]{mehoque@cs.rochester.edu}

\keywords{Self-organizing Social Networks $|$ Creativity $|$ Social Influence}

\begin{abstract}
Recent works suggest that striking a balance between maximizing idea stimulation and minimizing idea redundancy can elevate creativity in self-organizing social networks. We explore whether dispersing the visibility of idea generators can help achieve such a trade-off. We employ popularity signals (follower counts) of participants as an external source of variation in network structures, which we control across four randomized study conditions. We observe that popularity signals influence inspiration-seeking ties, partly by biasing people's perception of their peers' creativity. Networks that partially disperse the ideators' visibility using this external signal show reduced idea-redundancy and elevated creativity. However, extreme dispersal leads to inferior creativity by narrowing the range of idea stimulation. Our work holds future-of-work implications for elevating creativity.
\end{abstract}
\begin{document}

\flushbottom
\maketitle

\thispagestyle{empty}

\section*{Introduction}
Creative ideas are generated by individuals but creativity rarely emerges from a social vacuum.  What are the characteristics of social systems that maximize creativity? Consider two examples.

\textit{Example 1.} Idea generation in academia hardly ever happens in isolation. Researchers collaborate with and take inspiration from each other, thus creating a self-organizing, interconnected \textit{social system} of creative idea generators. Yet, the distribution of publications, citations, and funding at an individual level is extremely skewed~\cite{nielsen2021global}, allowing the `academic superstars' to outshine others in terms of visibility and contribution. This small fraction of superstars can be an irreplaceable source of ideas to their peers~\cite{azoulay2010superstar}. 

\textit{Example 2.} Creative art also rarely happens in isolation, and people more often than not build on top of others' creations. Superstars in an artistic domain can similarly influence their peers with skewed visibility and contribution. For instance, Nobel laureate Bob Dylan's creations influenced an entire generation of musicians, particularly in the folk genre.

The examples illustrate that despite being an individual strength, creativity benefits strongly from the \textit{social} stimulation of ideas. Our social embeddings exert a strong influence on the creative inspirations we find, just as it does on the information we receive, the decisions we make, and even the beliefs we hold~\cite{Almaatouq11379,shafipour2018buildup}. Question is, in a self-organizing social system, what kind of connection structure can best support the aggregate collection of creative ideas?

Recent explorations suggest two opposing factors that influence creativity in self-organizing social systems~\cite{baten2020creativity,baten2021cues}. On the one hand, people are drawn to seek creative inspirations from (or `follow', as referred to henceforth) the highly creative peers in their network. This positively impacts their creative performances, as highly creative ideas can help stimulate further novel ideas in people. On the other hand, this process also makes the high performers increasingly visible and central in the network (e.g., superstars). Consequently, people's sources of creative inspiration gradually become overlapping. This narrowing of stimuli idea-space can ironically lead to \textit{redundant} ideas being generated---a rather counter-productive outcome.

To best support the aggregate creative outcomes, we want an increased number of people to be inspired by the highly creative ideators, but not to the point where redundancy starts to stifle creativity in the social system. Intuitively, dispersing visibility in social networks (or `decentralizing' the networks) holds promise in countering redundancies. On the downside, such dispersal can expose people to sub-par stimuli, hurting the social stimulation of ideas in the first place. Here, we test the hypothesis that \textit{partially} dispersing in-degree centrality in social networks can help strike a balance between the two factors, and lead to superior creative outcomes.

\section*{Experimental Setup}
There are several challenges inherent in this line of inquiry, e.g., in identifying (1) what causes the network connection patterns to change across time (self-organization in social networks can be driven by many cues, such as skill/competence, success, prestige, and self-similarity~\cite{henrich2016secret}), (2) where people get their creative inspirations from (real-life data rarely allows for an unambiguous tracing of links between people's ideas and the respective stimuli), and (3) whether the network connection patterns have any impact on the creative outcomes.

To address these, we adopt an experimental paradigm where people's agency for self-organizing is retained, yet controls are built-in to allow us to definitively extract causal influences. We leverage popularity signals (e.g., follower counts) of the participants as an external source of variation in the network structures. Popularity or visibility signals have previously been reported to socially influence cultural, economic, and academic networks~\cite{salganik2006experimental}, underpinned by social psychological and behavioral economic heuristics. We use popularity signals to bias self-organization differently across various study conditions, which allows us to systematically assess the effects of network decentralization on people's creative performances.

We use a custom web interface to conduct a randomized controlled experiment in the virtual laboratory. \textit{First,} we adopt a text-based creativity task with 5 rounds of idea generation. We randomly assign the participants either of two roles: (i) \textit{alters}, whose ideas are recorded first to be used as stimuli, and (ii) \textit{egos}, who take creative inspirations from the alters. In each round, the egos first generate ideas independently (turn-1), then see the alters' ideas and submit any inspired ideas (turn-2), and finally, rate the alters' ideas and follow/unfollow them. This setting allows us to explicitly track the creative inspiration links as the networks evolve. \textit{Second,} the ideas of the same alters are shown to the egos of 4 different conditions, where we only modify the alters' follower counts as shown to the egos. This creates multiple, parallel histories of network evolution across conditions. \textit{Third,} we repeat the process for four independent trials, which helps us leverage a rich variation in the data for deriving the insights.

\begin{figure*}
    \centering
    \includegraphics[width=1\linewidth]{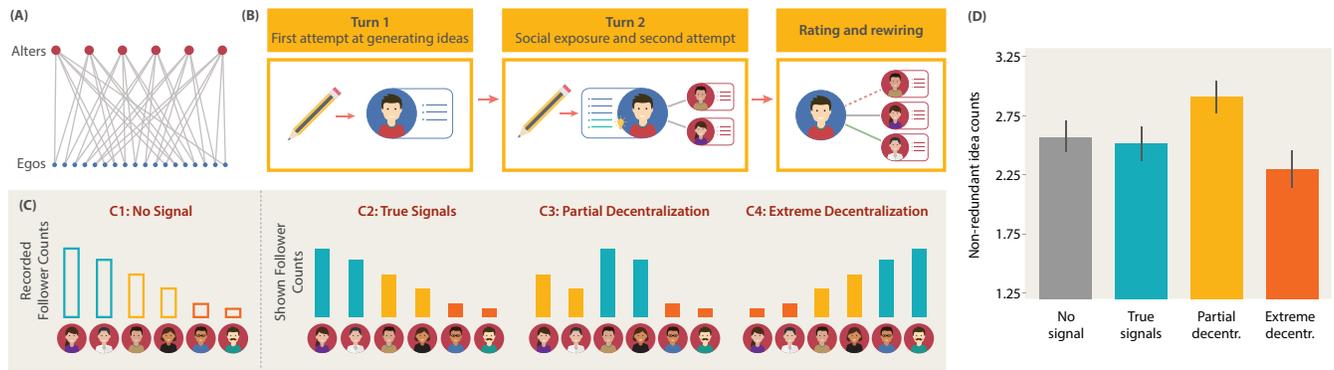}
    \caption{\textbf{\textit{(A)}} The initial bipartite network structure. Each ego is connected to $2$ alters out of $6$. \textbf{\textit{(B)}} Protocol for each of the $5$ rounds. In turn-1, an ego (blue) generates ideas independently. In turn-2, they view the ideas of the two alters (red) they are following, and submit any new ideas inspired by the stimuli. Then, they rate the ideas of all $6$ alters, and update which two alters to follow in the next round. \textbf{\textit{(C)}} The $4$ study conditions. The top, middle and bottom tiers of alters, as recorded from C1, are respectively shown in teal, mustard, and orange bars. \textbf{\textit{(D)}} Non-redundant idea counts across conditions. Whiskers denote $95\%$ C.I.}
    \label{infographic}
\end{figure*}

In the first condition (`C1: No signal'), no follower count is shown to the egos at all. From here we record the alters' true/unbiased follower counts in each round, and rank the alters into three tiers of popularity (top, middle, and bottom) based on their total follower counts in the 5 rounds. In the `C2: True signals' condition, we show the egos the true follower counts of the alters in each round, as recorded from C1. In the `C3: Partial decentralization' condition, we swap the follower counts of the first two tiers of alters, leaving the follower counts of the bottom tier alters untouched. This implicitly makes the second tier of alters appear more popular (and the first tier less popular) than they actually are in C1. In the `C4: Extreme decentralization' condition, the follower counts are shown in the reverse order of the alters' true popularity ranking. This condition makes the alters least popular in C1 appear the most popular, and vice versa, to bring even the least followed alters into the limelight (Figures~\ref{infographic}A-\ref{infographic}C).

\section*{Findings}
\subsection*{Popularity signals bias people's following patterns} People can conform to social influence when they assume others to be better informed than they are, or believe that the majority's pick is correct~\cite{deutsch1955study}. Here, the alters' popularity signals are the vehicles of social influence on the egos. Such signals can trigger anchoring biases, where a higher follower count shown on the screen can anchor an ego with an inflated impression of the alter's creativity, and vice versa~\cite{jacowitz1995measures}. We find that between pairs of conditions among C2 to C4, the change in an alter's shown follower counts had (i) a significant positive correlation with the change in their obtained follower counts (Pearson's $r$ = $0.52$, $P<10^{-4}$), and also (ii) a significant positive correlation with the change in the average ratings of their ideas (Pearson's $r$ = $0.11$, $P<0.04$). This suggests that the egos not only followed alters with inflated popularity signals more, but also perceived those alters' ideas to be more \textit{creative}. 

Since our network structure is bipartite (two-mode), we define a `decentralized network' to be one where every alter has an equal number of followers (same in-degree centrality)~\cite{freeman1978centrality}. We employ the Gini coefficient to capture the inequality in the alters' obtained follower counts (lower values denote higher decentralization). C4 showed a significantly lower Gini coefficient than every other condition (post-hoc 2-tailed tests, $P<0.05$ in all comparison cases), attesting that the obtained follower counts were spread out the most in this condition. In C3, the Gini coefficient among the first two tiers of alters was significantly less than the coefficient among all three tiers (2-tailed test, $P<0.05$; SI), showing that the decentralization happened partially among the first two tiers only, as intended. This provides validity to our choice of popularity signals as an external source of variation in the network structures.

\subsection*{Partial decentralization improves while extreme decentralization hurts creativity}
We employ two metrics of divergent creativity: (1) non-redundant idea count and (2) Creativity Quotient (higher is better in both; Methods). The C4 egos showed significantly lower performance than all other conditions in both of the metrics, while the C3 egos showed significantly higher non-redundant idea counts than all other conditions (Figure~\ref{infographic}D). At a collective-level, the C3 egos significantly outperformed other conditions in the total number of distinct ideas generated, while the C4 egos showed significantly lower collective-level creativity quotient than all other conditions (post-hoc Wilcoxon Rank Sum tests, $P<0.05$ in all comparison cases; SI). In summary, we observe that C3 improves while C4 hurts creative performances.

\subsection*{Idea redundancy reduces in partially decentralized networks} 
We shed light on the (1) ego-alter and (2) ego-ego overlaps of ideas, as measured by the Jaccard Index. We find that a lower overlap between the alters' ideas and the egos' turn-1 ideas is associated with an increased creative output of the egos in turn-2 (Creativity Quotient: Pearson's $r=-0.32$, $P<0.01$; Non-redundant idea count: $r=-0.11$, \textit{n.s.}). The associative theory of creative cognition suggests that stimuli ideas initially-unthought-of by a person can help them access remote concepts in their long-term memory and recombine those concepts to generate novel ideas~\cite{paulus2007toward}. Thus, a reduced overlap between the egos' and alters' ideas can better inspire novel recombinations in the egos, as we confirm. Following highly creative alters is known to reduce such ego-alter overlaps~\cite{baten2020creativity}. We also find that a reduced overlap/redundancy between turn-2 ideas of ego-pairs corresponds to increased performances by the egos (Non-redundant idea count: Pearson's $r=-0.55$, $P<10^{-6}$; Creativity Quotient: $r=-0.20$, $P=0.08$). The egos' similar choices of alters can exacerbate such idea redundancy~\cite{baten2020creativity}.

Importantly, C3 showed the lowest ego-ego redundancy among all conditions. This can help explain why C3 had elevated creative performances: by reducing redundancy while retaining sufficient creative stimulation. Interestingly, despite having the most spread-out visibility of the alters, C4 showed the \textit{highest} ego-ego redundancy. This can be explained by the fact that C4 brought the least performing alters (i.e., sub-par stimuli) to the fore, which can result in the generation of common ideas by the egos---leading to both increased ego-ego overlap and reduced creative performance as noted in the data.

\section*{Discussion}
We report that popularity signals can bias creative networks, partly by biasing people’s perception of their peers’ creativity. We find that a `Goldilocks' (not too hot, not too cold) extent of partial dispersal of visibility can help elevate creativity by reducing idea redundancies, whereas networks whose dispersal is too narrow or too broad tend to lead to inferior outcomes.

These results can have implications not only for the scientific study of creative networks~\cite{perry2003social}, but also for informing actionable insights in practical problems. For instance, research funding is increasingly being concentrated in the hands of fewer elites, and many call for spreading out resources to a large pool of recipients to ensure equity. We show that a middle ground can instead be the most rewarding. People increasingly seek creative inspiration from peers in online platforms like Behance, Pinterest, or ResearchGate. Our results show that content reception in such platforms may not remain purely meritocratic, but be socially influenced by the right-skewed popularity signals (e.g., upvotes). To counter redundancy, it can be useful to partially spread out which ideators are brought to the fore or recommended in such platforms. Due to the COVID-19 pandemic, many teams are brainstorming using remote collaboration platforms (e.g., Zoom, Slack). Our insights can inform AI-driven intervention strategies for elevating creative performances in such platforms. These will leave implications for the future-of-work, where people will increasingly need to perform creatively in a collaborative setting~\cite{frank2019toward,manyika2017future,baten2019upskilling}. 

It remains to be understood how to determine the optimal level of partial decentralization in a real-world, bidirectional, large-scale network, and how best to realize such decentralization practically. These are part of our future work directions.

\section*{Methods}
This study was approved by the IRB of the University of Rochester, USA. All participants provided informed consent.

\subsection*{Participants} $312$ participants were recruited from Amazon Mechanical Turk. Each of the 4 trials had $6$ alters and $72$ egos. The $72$ egos were split equally into the four conditions ($18$ egos in each).

\subsection*{Measures} Our creativity task is based on the canonical Alternate Uses Test~\cite{guildford1978alternate}. In each of the 5 rounds, an everyday object (e.g., a brick) was given along with its common use (e.g., used for building). The participants needed to generate alternative use ideas that were novel and useful, different from each other, different from the specified common use, appropriate, and feasible. The text-based ideas were scored using two metrics (see SI for details):

\subsubsection*{(i) Non-redundant idea counts} We binned the differently phrased yet same ideas together under common bin IDs. An idea was taken to be `non-redundant' if it was given by at most a threshold number of people. This metric quantifies how rare one's ideas are compared to their peers' ideas, but does not assess the ideas' intrinsic qualities.

\subsubsection*{(ii) Creativity Quotient} This metric uses information-theoretic methods to capture the semantic diversity of a person's idea-set~\cite{bossomaier2009semantic}. The intuition is, a diverse set of ideas likely touches many semantic categories, marking better creativity. However, the metric does not compare idea-sets socially like the first metric.

\subsection*{Procedure} Initially, we randomly assigned each ego to follow $2$ alters out of the $6$ in the trial, using the network structure shown in Figure~\ref{infographic}(A). Each of the $5$ rounds had three steps. First, the egos generated ideas independently (turn-1). Second, the egos were shown the ideas of the $2$ alters they were following. They could submit any new ideas that were inspired by the alters' ideas (turn-2). Third, the egos were shown the ideas of all $6$ alters, which they rated on novelty. The egos could optionally update which $2$ alters to follow in the next round (Figure~\ref{infographic}(B)). Only the alters' follower counts were shown differently across conditions. Each `tier' (ranked in C1) had two alters (Figure~\ref{infographic}(C)). See details in SI.

\subsection*{Capturing decentralization of the alters' creative influence} The popularity of an alter $i$ is defined by his/her share of followers, $m_i = d_i/ \sum_{k=1}^{S}d_k$, where $d_i$ is alter $i$’s follower count in a given round and $S$ is the number of alters in the trial. The Gini coefficient is given by, $G = \frac{\sum_{i=1}^S \sum_{j=1}^S |m_i - m_j| }{2S \sum_{k=1}^S m_k}$. This represents the average difference in follower counts between pairs of alters, normalized to fall between $0$ (complete equality/decentralization) and $1$ (maximum inequality/centralization).

\subsection*{Capturing overlaps between idea-sets} The Jaccard Index captures the overlap between two idea-sets, $A$ and $B$, as, $J(A,B)=|A\cap B| / |A\cup B|$. If $A=B=\emptyset$, we take $J(A,B)=1$. The idea-sets consist of bin IDs from the non-redundant idea count calculations.

\section*{Data Availability}
Please see \texttt{https://github.com/ROC-HCI/goldilocks-creativity-networks} for the data and code. Due to the copyright protection of the creativity test, we provide processed data of the participants' ideas.


\section*{Acknowledgements}

This work was supported by funding from the National Science Foundation (IIS-1750380), US Army Research Office (W911NF-18-1-0421), and National Institutes of Health (HD-037082).

\section*{Author contributions statement}

R.A.B. designed the study, collected and analyzed the data, interpreted the results, and authored the manuscript. R.N.A., G.G., and E.H. oversaw the study
design, interpretation of results, and preparation of the manuscript.

\section*{Additional information}
\textbf{Competing interests} The authors declare no competing interest.

\renewcommand{\thefigure}{S\arabic{figure}}
\renewcommand{\thetable}{S\arabic{table}}
\setcounter{figure}{0}
\setcounter{table}{0}

\clearpage
\date{}


\part*{Supplementary Information}
\textbf{\large Creativity in social networks is enhanced by `Goldilocks' dispersal of ideators' visibility}\\

\noindent \author{Raiyan Abdul Baten, Richard N. Aslin, Gourab Ghoshal, and Ehsan Hoque}\\

\noindent Corresponding Author: Ehsan Hoque,  mehoque@cs.rochester.edu


\maketitle


\clearpage
\section*{Extended Materials and Methods}

\subsection*{Participants}
All of the participants were located in the United States. They were assigned their roles (alter/ego) and study conditions randomly. Among the $312$ participants, $121$, $189$ and $2$ participants respectively self-identified to be of female, male and other gender. The racial distribution was: White: $238$, Black or African American: $33$, Asian: $21$, Two or More Races: $5$, other: $15$. $15$ participants belonged to Hispanic or Latino ethnicity. The age distribution was: 18-24: $24$, 25-34: $136$, 35-44: $82$, 45-54: $45$, 55+: $25$. Each of the 4 trials had $6$ alters and $72$ egos. The $72$ egos were split equally into the four conditions ($18$ egos in each).

\subsection*{Measures}
Unlike convergent thinking, which requires individuals to zero in on known correct answers (as tested in traditional school exams), divergent thinking leads people to generate \textit{numerous} and \textit{varied} responses to a given prompt or situation~\cite{runco2014creativity,kozbelt2010theories}. Our task is based on Guilford's Alternate Uses Test\footnote{(Guilford's Alternate Uses Test is Copyright @ 1960 by Sheridan Supply Co., all rights reserved in all media, and is published by Mind Garden, Inc, www.mindgarden.com)}~\cite{guildford1978alternate}, the canonical approach for quantifying divergent creative performances. We chose the first $5$ objects from Form B of Guilford's test as the prompt objects in the $5$ rounds. In each round, the participants were given an everyday object (e.g., a brick), whose common use was specified (e.g., a brick is used for building). The participants needed to come up with alternative use ideas for the object: ideas that were novel and useful, different from each other, different than the specified common use, appropriate, and feasible. The participants were guided use examples throughout the study, as specified in the test manual.

We used two complementary metrics for quantifying creativity. The Non-redundant Idea Count metric quantifies how rare one's ideas are compared to their peers' ideas~\cite{oppezzo2014give,abdullah2016shining}, and does not capture the idea qualities per se (even a great idea is not rare/creative if many people submit it). In contrast, Creativity Quotient captures how semantically diverse a person's idea-set is~\cite{snyder2004creativity,bossomaier2009semantic}, and does not attempt to compare the idea-set socially (two people having highly diverse, yet identical, idea-sets will achieve identically high scores). Further details are given below:

\subsubsection*{Non-redundant Idea Count} 
We first discarded inappropriate submissions that did not meet the specified requirements. Since the same idea can be phrased differently by different people, we collected or `binned' the same ideas together under common bin IDs. For binning the ideas, we followed the coding rules described by Bouchard and Hare~\cite{bouchard1970size} and the scoring key of Guilford's test. Based on the bin IDs, an idea was marked to be non-redundant if it was given by at most a threshold number of participants in a given pool of ideas. For the alters, the threshold was set to $1$, and the pools were the round-wise idea-sets of the $6$ alters in the trial. For the egos, the threshold was set to $2$, and the pools comprised of the round-wise ideas of the egos in a given study condition in a given trial.

The first author binned all of the ideas in the dataset, while two other research assistants independently binned the ideas of a random $25\%$ of the participants. They were shown the ideas in a random order. Based on their independent bin IDs, the total non-redundant idea counts of the participants in all $5$ rounds were computed, and the agreements between the coders were calculated. The agreements were high both between the first and second coder (intra-class correlation coefficient, $ICC(3,2)=0.93$, $P<10^{-13}$, $95\%$ C.I. = $[0.88,0.96]$, Pearson's $r=0.89$, $P<10^{-14}$, $95\%$ C.I.=$[0.81,0.94]$), and between the first and third coder (intra-class correlation coefficient $ICC(3,2)=0.87$, $P<10^{-11}$, $95\%$ C.I. = $[0.80,0.92]$, Pearson's $r=0.83$, $P<10^{-13}$, $95\%$ C.I.=$[0.72,0.90]$). We used the bin annotations from the first coder in the analyses. 

For collective-level analysis, we took the total number of distinct bin IDs generated by the egos in a given condition in a given trial as the collective non-redundancy marker.

\subsubsection*{Creativity Quotient}


If the ideas of a person are very similar to each other, they are likely subtle or incremental variations of a small number of semantic categories. Contrarily, if the ideas are mutually dissimilar, they likely touched many semantic categories, marking better creativity~\cite{rietzschel2007personal}. Creativity Quotient, $Q$, captures this intuition computationally~\cite{snyder2004creativity,bossomaier2009semantic}, by employing an information theoretic measure of semantic similarity derived from WordNet~\cite{miller1995wordnet}. 

Concepts are organized as syn-sets or synonym sets in WordNet, where the nouns are linked with `is a' relationships. We removed stop-words and punctuation from the ideas before running a spell-checker on them. Next, we split the ideas into their constituting set of concepts, and converted the terms into nouns whenever possible. We then computed the information content of those concepts. The taxonomic organization of WordNet implies that concepts with many hyponyms convey less information than concepts with less number of hyponyms~\cite{seco2004intrinsic}. Therefore, infrequent concepts at the leaf nodes hold more information than their abstracting nodes. The Information Content, $I$, of a concept $c$ can thus be calculated as,
\begin{equation}
    I(c) = \frac{log\big(\frac{h(c)+1}{w}\big)}{log\big(\frac{1}{w}\big)} = 1- \frac{log(h(c)+1)}{log(w)},
\end{equation}
where $h(c)$ is the number of hyponyms of $c$, and $w$ is the total number of concepts in WordNet. The denominator ensures $I\in[0,1]$ by normalizing the metric against the most informative concept.

Given a participant's set of ideas in turn-2 of a given round, we proceeded to compute the semantic diversity as follows. Between every pair of concepts in the idea-set, $c_1$ and $c_2$, we calculated the semantic similarity~\cite{jiang1997semantic} as,
\begin{equation}
    sim(c_1,c_2)= 1-\Big(\frac{I(c_1)+I(c_2)-2\times sim_{MSCA}(c_1,c_2)}{2} \Big),
\end{equation}
where $sim(c_1,c_2)$ is a function of the information overlap between the two concepts, $sim_{MSCA}(c_1,c_2)$. This overlap, in turn, is computed using the information content of the Most Specific Common Abstraction (MSCA) that subsumes both of the concepts,
\begin{equation}
    sim_{MSCA}(c_1,c_2) = \max_{c'\in S(c_1,c_2)} I(c'),
\end{equation}
where $S(c_1,c_2)$ is the set of concepts subsuming $c_1$ and $c_2$.

Given the pair-wise concept similarities, we computed the multi-information, $I_m$, as the shared information across the idea-set. We crafted the max spanning tree from the network of concepts and their pairwise similarity values. We summed over the edge weights in the max spanning tree to get $I_m$. Finally, we calculated $Q$ as,
\begin{equation}
    Q = N-I_m,
\end{equation}
where $N$ is the total number of concepts in the person's idea-set. 

To compute collective-level Creativity Quotients, we collected in a bag-of-words document all of the ideas generated in turn-2 of each round by all of the trial-wise egos, and calculated the Creativity Quotient of the document using the same procedure as above.

\subsection*{Procedure}
The egos were given $3$ minutes in each of turn-1 and turn-2 to generate ideas on the given object. They were forbidden to re-submit their followee alters' ideas exactly, and were told that their non-redundant idea counts would contribute to their creative performances. The egos were also informed of a short test to be held at the end of the study, where they would need to recall ideas shown to them. This was in place to ensure the participants' attention to the stimuli ideas, which is known to have positive stimulation benefits~\cite{nijstad2006group,dugosh2005cognitive,paulus2000groups,brown1998modeling}. After generating ideas for two turns, the egos rated the ideas of all of the $6$ alters in their trial on a 5-point Likert scale (1: not novel, 5: highly novel). As the egos (optionally) followed/unfollowed alters in each round, they were required to submit the rationale behind updating/not updating their links. This helped ensure the accountability of the egos in making their choices, which is known to raise epistemic motivation and improve systematic information processing~\cite{bechtoldt2010motivated,scholten2007motivated}. The participants were paid \$10 upon the completion of the tasks, as well a bonus of \$5 if they were among the top 5 performers in their trials.

\section*{Extended Statistical Analysis Results}
\subsection*{Popularity signals bias people's following patterns}
\subsubsection*{Correlation analysis}
Between pairs of conditions among C2 to C4, we computed the differences in the (i) shown popularity signals ($\Delta p_{shown}$), (ii) obtained follower counts ($\Delta p_{obtained}$), and (iii) obtained average ratings of the ideas ($\Delta r_{obtained}$) of each alter in each round.

We found that the change in an alter's shown follower counts, $\Delta p_{shown}$, had (i) a significant positive correlation with the change in their obtained follower counts, $\Delta p_{obtained}$ (Pearson's $r$ = $0.52$, $P<10^{-4}$, $95\%$ C.I.=[$0.44$, $0.59$]), and also (ii) a significant positive correlation with the change in the average ratings of their ideas, $\Delta r_{obtained}$ (Pearson's $r$ = $0.11$, $P<0.04$, $95\%$ C.I.=[$0.004$, $0.21$]). This suggests that the egos not only followed alters with inflated popularity signals more, but also perceived those alters' ideas to be more \textit{creative}. 

\subsubsection*{Gini coefficient analysis}
Since our network structure is bipartite, we quantified network decentralization based on how similar the follower counts of the alters were in a given trial. For example, if all of the alters obtained an equal number of followers, then the network would be perfectly decentralized and the egos' inspiration sources would be maximally spread out. To this end, we employed the Gini Coefficient, which is a measure of success (popularity) inequality, to capture network decentralization. A lower value of the coefficient corresponds to a higher level of decentralization. 

To capture partial versus extreme decentralization, we computed the Gini coefficient separately among (A) the top 2 tiers of alters (out of 3 tiers), and (B) all 3 tiers of alters. For example, if the top 2 tiers of alters are significantly more decentralized than the full set of alters, then the network is only partially decentralized. We ran the statistical tests using a $4 \times 2$ factorial design, with $4$ levels in the `Condition' factor (C1 to C4), and $2$ levels in the `Tiers' factor (top $2$ tiers, and all $3$ tiers). We analyzed the data using Linear Mixed Models (LMM). Since the study consisted of $5$ rounds of creative ideation with $5$ different objects, we assessed the fixed effects of the two factors against the random effects of the Round IDs to account for repeated measures.

We found a significant main effect of the `Condition' factor ($F(3,148)=4.8$, $P<0.004$), and a marginal main effect of the `Tiers' factor ($F(1,148)=3.71$, $P=0.056$). The interaction between the two factors was insignificant. Post-hoc pairwise comparisons showed that among all 3 tiers of alters, C4 indeed showed significantly less Gini Coefficient than each of the other three conditions ($2$-tailed tests, C1 vs C4: $t(148)=4.35$, $P<0.001$; C2 vs C4: $t(148)=3.72$, $P<0.01$; C3 vs C4: $t(148)=3.22$, $P<0.04$). Only in C3, the Gini coefficient among the first two tiers of alters was significantly less than the coefficient among all three tiers ($2$-tailed test, $t(148)=-3.15$, $P=0.037$)---confirming the partial decentralization. All of the $P$ values are corrected for multiple comparisons using Holm's Sequential Bonferroni Procedure.

\subsection*{Partial decentralization improves while complete decentralization hurts creativity} We collected the turn-2 non-redundant idea counts of the egos from different study conditions. Using the non-parametric Kruskal-Wallis test, we found a significant difference across conditions in the omnibus test ($\chi^2$(df$=3$, $N=1440$)=$38.93$, $P<10^{-7}$). Post-hoc pairwise comparisons using the Wilcoxon Rank Sum test showed that C3 scored significantly higher than each of the other three conditions (C1 vs C3: $W = 55911$, $P = 0.004$; C2 vs C3: $W = 54298$, $P<0.001$; C4 vs C3: $W = 81338$, $P<0.001$), while C4 scored significantly lower than each of the other conditions (C1 vs C4: $W = 73250$, $P = 0.006$; C2 vs C4: $W = 71020$, $P=0.045$). 

We repeated the same analysis for the egos' Creativity Quotients in turn-2 in each round. The Kruskal-Wallis test showed a significant difference across conditions in the omnibus test ($\chi^2$(df$=3$, $N=1440$)=$14.57$, $P<0.01$). Post-hoc pairwise comparisons using the Wilcoxon Rank Sum test showed that C4 scored significantly lower than each of the other conditions (C1 vs C4: $W = 73904$, $P = 0.006$; C2 vs C4: $W = 72323$, $P=0.028$; C3 vs C4: $W = 73996$, $P=0.006$). 

As for collective-level outcomes, the Kruskal-Wallis test showed a significant difference across conditions in the omnibus test ($\chi^2$(df$=3$, $N=80$)=$25.48$, $P<10^{-4}$). Post-hoc pairwise comparisons using the Wilcoxon Rank Sum test showed that C3 outperformed each of the other three conditions (C1 vs C3: $W = 57.5$, $P = 0.001$; C2 vs C3: $W = 96.5$, $P=0.021$; C4 vs C3: $W = 360$, $P<0.001$). C4 performed marginally lower than C2 ($W=287.5$, $P=0.055$).

Next, we computed collective-level Creativity Quotients. Namely, we collected in a bag-of-words document all of the ideas generated in turn-2 of each round by all of the trial-wise egos in a given condition, and computed the Creativity Quotient of the document. The Kruskal-Wallis test showed a significant difference across conditions in the omnibus test ($\chi^2$(df$=3$, $N=80$)=$19.58$, $P<10^{-3}$). Post-hoc pairwise comparisons using the Wilcoxon Rank Sum test showed that C4 performed lower than each of the other three conditions (C1 vs C4: $W = 340$, $P = 0.001$; C2 vs C4: $W = 315$, $P=0.008$; C3 vs C4: $W = 337$, $P=0.001$).

In summary, we observed that C3 improved while C4 hurt creative performances. All of the $P$ values reported here are corrected for multiple comparisons using Holm's Sequential Bonferroni Procedure.

\subsection*{Idea redundancy reduces in partially decentralized networks} 

We employed the Jaccard Index to quantify the overlaps between pairs of idea-sets. As the ego-ego overlap decreased, the average non-redundant idea-count showed a significantly increasing trend (Pearson's $r=-0.55$, $P<10^{-6}$, $95\%$ C.I.=[$-0.68,-0.37$]), while the average creativity quotient showed a moderately increasing trend (Pearson's $r=-0.20$, $P=0.08$, $95\%$ C.I.=[$-0.40,0.03$]). Similarly, a decrease in the ego-alter overlap corresponded to a significantly increasing trend in the creativity quotient metric (Pearson's $r=-0.32$, $P<0.01$, $95\%$ C.I.=[$-0.50,-0.10$]), while the non-redundant idea count metric also showed a slightly increasing trend (Pearson's $r=-0.11$, $P>0.05$).

The omnibus Kruskal-Wallis test showed that the ego-ego overlaps varied significantly across conditions ($\chi^2$(df$=3$,$N=80$)=$16.88$, $P<0.001$). Post-hoc pairwise comparisons using the Wilcoxon Rank Sum test showed that C3 had lower ego-ego overlaps than all other conditions, with the differences with C1 and C4 being statistically significant (C1 vs C3: $W = 325$, $P = 0.005$; C2 vs C3: $W = 229$, $P=0.441$; C4 vs C3: $W = 80$, $P=0.006$). C4 showed the highest ego-ego overlaps among all conditions, with the differences with C3 being statistically significant and the differences with C2 being marginally significant (C2 vs C4: $W = 112$, $P=0.072$; C3 vs C4: $W = 80$, $P=0.006$). All of the $P$ values reported here are corrected for multiple comparisons using Holm's Sequential Bonferroni Procedure.


\begin{thebibliography}{10}
\urlstyle{rm}
\expandafter\ifx\csname url\endcsname\relax
  \def\url#1{\texttt{#1}}\fi
\expandafter\ifx\csname urlprefix\endcsname\relax\def\urlprefix{URL }\fi
\expandafter\ifx\csname doiprefix\endcsname\relax\def\doiprefix{DOI: }\fi
\providecommand{\bibinfo}[2]{#2}
\providecommand{\eprint}[2][]{\url{#2}}

\bibitem{nielsen2021global}
\bibinfo{author}{Nielsen, M.~W.} \& \bibinfo{author}{Andersen, J.~P.}
\newblock \bibinfo{journal}{\bibinfo{title}{Global citation inequality is on
  the rise}}.
\newblock {\emph{\JournalTitle{Proceedings of the National Academy of
  Sciences}}} \textbf{\bibinfo{volume}{118}} (\bibinfo{year}{2021}).

\bibitem{azoulay2010superstar}
\bibinfo{author}{Azoulay, P.}, \bibinfo{author}{Graff~Zivin, J.~S.} \&
  \bibinfo{author}{Wang, J.}
\newblock \bibinfo{journal}{\bibinfo{title}{Superstar extinction}}.
\newblock {\emph{\JournalTitle{The Quarterly Journal of Economics}}}
  \textbf{\bibinfo{volume}{125}}, \bibinfo{pages}{549--589}
  (\bibinfo{year}{2010}).

\bibitem{Almaatouq11379}
\bibinfo{author}{Almaatouq, A.} \emph{et~al.}
\newblock \bibinfo{journal}{\bibinfo{title}{Adaptive social networks promote
  the wisdom of crowds}}.
\newblock {\emph{\JournalTitle{Proceedings of the National Academy of
  Sciences}}} \textbf{\bibinfo{volume}{117}}, \bibinfo{pages}{11379--11386},
  \doiprefix\url{10.1073/pnas.1917687117} (\bibinfo{year}{2020}).
\newblock \eprint{https://www.pnas.org/content/117/21/11379.full.pdf}.

\bibitem{shafipour2018buildup}
\bibinfo{author}{Shafipour, R.} \emph{et~al.}
\newblock \bibinfo{journal}{\bibinfo{title}{Buildup of speaking skills in an
  online learning community: A network-analytic exploration}}.
\newblock {\emph{\JournalTitle{Palgrave Communications}}}
  \textbf{\bibinfo{volume}{4}}, \bibinfo{pages}{63} (\bibinfo{year}{2018}).

\bibitem{baten2020creativity}
\bibinfo{author}{Baten, R.~A.} \emph{et~al.}
\newblock \bibinfo{journal}{\bibinfo{title}{Creativity in temporal social
  networks: How divergent thinking is impacted by one’s choice of peers}}.
\newblock {\emph{\JournalTitle{Journal of the Royal Society Interface}}}
  \textbf{\bibinfo{volume}{17}}, \bibinfo{pages}{20200667}
  (\bibinfo{year}{2020}).

\bibitem{baten2021cues}
\bibinfo{author}{Baten, R.~A.}, \bibinfo{author}{Aslin, R.~N.},
  \bibinfo{author}{Ghoshal, G.} \& \bibinfo{author}{Hoque, E.}
\newblock \bibinfo{journal}{\bibinfo{title}{Cues to gender and racial identity
  reduce creativity in diverse social networks}}.
\newblock {\emph{\JournalTitle{Scientific Reports}}}
  \textbf{\bibinfo{volume}{11}}, \bibinfo{pages}{1--10} (\bibinfo{year}{2021}).

\bibitem{henrich2016secret}
\bibinfo{author}{Henrich, J.}
\newblock \emph{\bibinfo{title}{The Secret of our Success: How Culture is
  Driving Human Evolution, Domesticating our Species, and Making us Smarter}}
  (\bibinfo{publisher}{Princeton University Press}, \bibinfo{year}{2016}).

\bibitem{salganik2006experimental}
\bibinfo{author}{Salganik, M.~J.}, \bibinfo{author}{Dodds, P.~S.} \&
  \bibinfo{author}{Watts, D.~J.}
\newblock \bibinfo{journal}{\bibinfo{title}{Experimental study of inequality
  and unpredictability in an artificial cultural market}}.
\newblock {\emph{\JournalTitle{Science}}} \textbf{\bibinfo{volume}{311}},
  \bibinfo{pages}{854--856} (\bibinfo{year}{2006}).

\bibitem{deutsch1955study}
\bibinfo{author}{Deutsch, M.} \& \bibinfo{author}{Gerard, H.~B.}
\newblock \bibinfo{journal}{\bibinfo{title}{A study of normative and
  informational social influences upon individual judgment}}.
\newblock {\emph{\JournalTitle{The Journal of Abnormal and Social Psychology}}}
  \textbf{\bibinfo{volume}{51}}, \bibinfo{pages}{629} (\bibinfo{year}{1955}).

\bibitem{jacowitz1995measures}
\bibinfo{author}{Jacowitz, K.~E.} \& \bibinfo{author}{Kahneman, D.}
\newblock \bibinfo{journal}{\bibinfo{title}{Measures of anchoring in estimation
  tasks}}.
\newblock {\emph{\JournalTitle{Personality and Social Psychology Bulletin}}}
  \textbf{\bibinfo{volume}{21}}, \bibinfo{pages}{1161--1166}
  (\bibinfo{year}{1995}).

\bibitem{freeman1978centrality}
\bibinfo{author}{Freeman, L.~C.}
\newblock \bibinfo{journal}{\bibinfo{title}{Centrality in social networks
  conceptual clarification}}.
\newblock {\emph{\JournalTitle{Social Networks}}} \textbf{\bibinfo{volume}{1}},
  \bibinfo{pages}{215--239} (\bibinfo{year}{1978}).

\bibitem{paulus2007toward}
\bibinfo{author}{Paulus, P.~B.} \& \bibinfo{author}{Brown, V.~R.}
\newblock \bibinfo{journal}{\bibinfo{title}{Toward more creative and innovative
  group idea generation: A cognitive-social-motivational perspective of
  brainstorming}}.
\newblock {\emph{\JournalTitle{Social and Personality Psychology Compass}}}
  \textbf{\bibinfo{volume}{1}}, \bibinfo{pages}{248--265}
  (\bibinfo{year}{2007}).

\bibitem{perry2003social}
\bibinfo{author}{Perry-Smith, J.~E.} \& \bibinfo{author}{Shalley, C.~E.}
\newblock \bibinfo{journal}{\bibinfo{title}{The social side of creativity: A
  static and dynamic social network perspective}}.
\newblock {\emph{\JournalTitle{Academy of Management Review}}}
  \textbf{\bibinfo{volume}{28}}, \bibinfo{pages}{89--106}
  (\bibinfo{year}{2003}).

\bibitem{frank2019toward}
\bibinfo{author}{Frank, M.~R.} \emph{et~al.}
\newblock \bibinfo{journal}{\bibinfo{title}{Toward understanding the impact of
  artificial intelligence on labor}}.
\newblock {\emph{\JournalTitle{Proceedings of the National Academy of
  Sciences}}} \textbf{\bibinfo{volume}{116}}, \bibinfo{pages}{6531--6539}
  (\bibinfo{year}{2019}).

\bibitem{manyika2017future}
\bibinfo{author}{Manyika, J.} \emph{et~al.}
\newblock \bibinfo{title}{A future that works: Automation, employment, and
  productivity} (\bibinfo{year}{2017}).

\bibitem{baten2019upskilling}
\bibinfo{author}{Baten, R.~A.}, \bibinfo{author}{Clark, F.} \&
  \bibinfo{author}{Hoque, M.~E.}
\newblock \bibinfo{title}{Upskilling together: How peer-interaction influences
  speaking-skills development online}.
\newblock In \emph{\bibinfo{booktitle}{8th International Conference on
  Affective Computing and Intelligent Interaction (ACII)}},
  \bibinfo{pages}{662--668} (\bibinfo{organization}{IEEE},
  \bibinfo{year}{2019}).

\bibitem{guildford1978alternate}
\bibinfo{author}{Guilford, J.}, \bibinfo{author}{Christensen, P.},
  \bibinfo{author}{Merrifield, P.} \& \bibinfo{author}{Wilson, R.}
\newblock \emph{\bibinfo{title}{Alternate Uses: Manual of Instructions and
  Interpretation}}.
\newblock \bibinfo{organization}{Orange, CA: Sheridan Psychological Services}
  (\bibinfo{year}{1978}).

\bibitem{bossomaier2009semantic}
\bibinfo{author}{Bossomaier, T.}, \bibinfo{author}{Harr{\'e}, M.},
  \bibinfo{author}{Knittel, A.} \& \bibinfo{author}{Snyder, A.}
\newblock \bibinfo{journal}{\bibinfo{title}{A semantic network approach to the
  {C}reativity {Q}uotient ({CQ})}}.
\newblock {\emph{\JournalTitle{Creativity Research Journal}}}
  \textbf{\bibinfo{volume}{21}}, \bibinfo{pages}{64--71}
  (\bibinfo{year}{2009}).

\bibitem{runco2014creativity}
\bibinfo{author}{Runco, M.~A.}
\newblock \emph{\bibinfo{title}{Creativity: Theories and Themes: Research,
  Development, and Practice}} (\bibinfo{publisher}{Elsevier},
  \bibinfo{year}{2014}).

\bibitem{kozbelt2010theories}
\bibinfo{author}{Kozbelt, A.}, \bibinfo{author}{Beghetto, R.~A.} \&
  \bibinfo{author}{Runco, M.~A.}
\newblock \bibinfo{journal}{\bibinfo{title}{Theories of creativity}}.
\newblock {\emph{\JournalTitle{The Cambridge Handbook of Creativity}}}
  \textbf{\bibinfo{volume}{2}}, \bibinfo{pages}{20--47} (\bibinfo{year}{2010}).

\bibitem{oppezzo2014give}
\bibinfo{author}{Oppezzo, M.} \& \bibinfo{author}{Schwartz, D.~L.}
\newblock \bibinfo{journal}{\bibinfo{title}{Give your ideas some legs: The
  positive effect of walking on creative thinking.}}
\newblock {\emph{\JournalTitle{Journal of Experimental Psychology: Learning,
  Memory, and Cognition}}} \textbf{\bibinfo{volume}{40}}, \bibinfo{pages}{1142}
  (\bibinfo{year}{2014}).

\bibitem{abdullah2016shining}
\bibinfo{author}{Abdullah, S.}, \bibinfo{author}{Czerwinski, M.},
  \bibinfo{author}{Mark, G.} \& \bibinfo{author}{Johns, P.}
\newblock \bibinfo{title}{Shining (blue) light on creative ability}.
\newblock In \emph{\bibinfo{booktitle}{Proceedings of the 2016 ACM
  International Joint Conference on Pervasive and Ubiquitous Computing}},
  \bibinfo{pages}{793--804} (\bibinfo{organization}{ACM},
  \bibinfo{year}{2016}).

\bibitem{snyder2004creativity}
\bibinfo{author}{Snyder, A.}, \bibinfo{author}{Mitchell, J.},
  \bibinfo{author}{Bossomaier, T.} \& \bibinfo{author}{Pallier, G.}
\newblock \bibinfo{journal}{\bibinfo{title}{The {C}reativity {Q}uotient: An
  objective scoring of ideational fluency}}.
\newblock {\emph{\JournalTitle{Creativity Research Journal}}}
  \textbf{\bibinfo{volume}{16}}, \bibinfo{pages}{415--419}
  (\bibinfo{year}{2004}).

\bibitem{bouchard1970size}
\bibinfo{author}{Bouchard~Jr, T.~J.} \& \bibinfo{author}{Hare, M.}
\newblock \bibinfo{journal}{\bibinfo{title}{Size, performance, and potential in
  brainstorming groups}}.
\newblock {\emph{\JournalTitle{Journal of Applied Psychology}}}
  \textbf{\bibinfo{volume}{54}}, \bibinfo{pages}{51} (\bibinfo{year}{1970}).

\bibitem{rietzschel2007personal}
\bibinfo{author}{Rietzschel, E.~F.}, \bibinfo{author}{De~Dreu, C.~K.} \&
  \bibinfo{author}{Nijstad, B.~A.}
\newblock \bibinfo{journal}{\bibinfo{title}{Personal need for structure and
  creative performance: The moderating influence of fear of invalidity}}.
\newblock {\emph{\JournalTitle{Personality and Social Psychology Bulletin}}}
  \textbf{\bibinfo{volume}{33}}, \bibinfo{pages}{855--866}
  (\bibinfo{year}{2007}).

\bibitem{miller1995wordnet}
\bibinfo{author}{Miller, G.~A.}
\newblock \bibinfo{journal}{\bibinfo{title}{Word{N}et: A lexical database for
  {E}nglish}}.
\newblock {\emph{\JournalTitle{Communications of the ACM}}}
  \textbf{\bibinfo{volume}{38}}, \bibinfo{pages}{39--41}
  (\bibinfo{year}{1995}).

\bibitem{seco2004intrinsic}
\bibinfo{author}{Seco, N.}, \bibinfo{author}{Veale, T.} \&
  \bibinfo{author}{Hayes, J.}
\newblock \bibinfo{title}{An intrinsic information content metric for semantic
  similarity in {W}ord{N}et}.
\newblock In \emph{\bibinfo{booktitle}{Proceedings of the 16th Eureopean
  Conference on Artificial Intelligence, ECAI}}, vol.~\bibinfo{volume}{16},
  \bibinfo{pages}{1089} (\bibinfo{year}{2004}).

\bibitem{jiang1997semantic}
\bibinfo{author}{Jiang, J.~J.} \& \bibinfo{author}{Conrath, D.~W.}
\newblock \bibinfo{title}{Semantic similarity based on corpus statistics and
  lexical taxonomy}.
\newblock In \emph{\bibinfo{booktitle}{Proceedings of the International
  Conference on Research in Computational Linguistics}} (\bibinfo{year}{1998}).

\bibitem{nijstad2006group}
\bibinfo{author}{Nijstad, B.~A.} \& \bibinfo{author}{Stroebe, W.}
\newblock \bibinfo{journal}{\bibinfo{title}{How the group affects the mind: A
  cognitive model of idea generation in groups}}.
\newblock {\emph{\JournalTitle{Personality and Social Psychology Review}}}
  \textbf{\bibinfo{volume}{10}}, \bibinfo{pages}{186--213}
  (\bibinfo{year}{2006}).

\bibitem{dugosh2005cognitive}
\bibinfo{author}{Dugosh, K.~L.} \& \bibinfo{author}{Paulus, P.~B.}
\newblock \bibinfo{journal}{\bibinfo{title}{Cognitive and social comparison
  processes in brainstorming}}.
\newblock {\emph{\JournalTitle{Journal of Experimental Social Psychology}}}
  \textbf{\bibinfo{volume}{41}}, \bibinfo{pages}{313--320}
  (\bibinfo{year}{2005}).

\bibitem{paulus2000groups}
\bibinfo{author}{Paulus, P.}
\newblock \bibinfo{journal}{\bibinfo{title}{Groups, teams, and creativity: The
  creative potential of idea-generating groups}}.
\newblock {\emph{\JournalTitle{Applied Psychology}}}
  \textbf{\bibinfo{volume}{49}}, \bibinfo{pages}{237--262}
  (\bibinfo{year}{2000}).

\bibitem{brown1998modeling}
\bibinfo{author}{Brown, V.}, \bibinfo{author}{Tumeo, M.},
  \bibinfo{author}{Larey, T.~S.} \& \bibinfo{author}{Paulus, P.~B.}
\newblock \bibinfo{journal}{\bibinfo{title}{Modeling cognitive interactions
  during group brainstorming}}.
\newblock {\emph{\JournalTitle{Small Group Research}}}
  \textbf{\bibinfo{volume}{29}}, \bibinfo{pages}{495--526}
  (\bibinfo{year}{1998}).

\bibitem{bechtoldt2010motivated}
\bibinfo{author}{Bechtoldt, M.~N.}, \bibinfo{author}{De~Dreu, C.~K.},
  \bibinfo{author}{Nijstad, B.~A.} \& \bibinfo{author}{Choi, H.-S.}
\newblock \bibinfo{journal}{\bibinfo{title}{Motivated information processing,
  social tuning, and group creativity}}.
\newblock {\emph{\JournalTitle{Journal of Personality and Social Psychology}}}
  \textbf{\bibinfo{volume}{99}}, \bibinfo{pages}{622} (\bibinfo{year}{2010}).

\bibitem{scholten2007motivated}
\bibinfo{author}{Scholten, L.}, \bibinfo{author}{Van~Knippenberg, D.},
  \bibinfo{author}{Nijstad, B.~A.} \& \bibinfo{author}{De~Dreu, C.~K.}
\newblock \bibinfo{journal}{\bibinfo{title}{Motivated information processing
  and group decision-making: Effects of process accountability on information
  processing and decision quality}}.
\newblock {\emph{\JournalTitle{Journal of Experimental Social Psychology}}}
  \textbf{\bibinfo{volume}{43}}, \bibinfo{pages}{539--552}
  (\bibinfo{year}{2007}).

\end{thebibliography}
\end{document}